\documentclass{aa} 
\usepackage{epsfig}
\usepackage{graphicx}

\begin{document}

\title{Narrow band survey for intragroup light in the Leo HI
cloud\fnmsep\thanks{Based on observations carried out at the ESO MPI
2.2m telescope, La Silla, and at the VLT - UT4, Paranal, Chile.}}

\subtitle{Constraints on the galaxy background contamination in
imaging surveys for intracluster planetary nebulae}
 
\author{Nieves Castro-Rodr\'{\i}guez\inst{1,2} \and J. Alfonso
L. Aguerri\inst{1,2} \and Magda Arnaboldi\inst{3,4} \and Ortwin
Gerhard\inst{1} \and Kenneth C. Freeman\inst{5} \and Nicola R.
Napolitano\inst{4,6} \and Massimo Capaccioli\inst{4}}

\institute{Astronomisches Institut, Universitaet Basel, Venusstrasse
7, CH-4102 Binningen, Switzerland \and Instituto de Astrof\'{\i}sica
de Canarias, E-38205 La Laguna, Spain \and I.N.A.F., Osservatorio
Astronomico di Torino, Via Osservatorio 20, I-10025 Pino Torinese \and
I.N.A.F.,Osservatorio Astronomico di Capodimonte, Via Moiariello 16,
I-80131 Naples \and RSAA, Mt.\ Stromlo and Siding Spr.\ Obs.,
Private Bag, Woden P.\ O., Canberra, ACT 2606, Australia \and Kapteyn
Institute, Postbus 800, Groningen 9700 AV, Netherlands}

\offprints{ncastro@astro.unibas.ch}

\date{Received....; accepted......}


\abstract{We have observed emission line objects located in a 0.26
  deg$^2$ field in the M96 (Leo) group, coincident with the
  intergalactic HI cloud. The emission line objects were selected
  using the same procedure as used for the search for intracluster
  planetary nebulae in the Virgo cluster. 29 emission line objects
  were identified, with [OIII] filter magnitudes brighter than
  $m_{5007}=28$.  Their $m_{5007}$ luminosity function has a bright
  cut-off $\simeq 1.2$ magnitude fainter than for the luminosity
  function of the planetary nebulae (PN) associated with the
  elliptical galaxies in the M96 group.  Therefore { the vast
  majority of these emission line objects are compatible with not
  being intragroup planetary nebulae at the Leo group distance of 10
  Mpc}.  Spectroscopic follow-up of two emission line objects in this
  Leo field showed that indeed these do not have the [OIII] $\lambda
  4959,\, \lambda 5007$ \AA\ doublet expected for a real PN.  The
  brighter source is identified as a starburst object at redshift $z =
  3.128$, because of a second emission in the near infrared,
  identified as FeII ($\lambda 2220$ \AA).
  
  From these data we derive three main results: (i) from the absence
  of PN we can determine a more stringent upper limit to the surface
  brightness in any old stellar population associated with the Leo HI
  cloud { at the surveyed position}, $\mu_{B,*}<32.8\, {\rm mag\,
  arcsec}^{- 2}$. (ii) This translates to an upper limit of $1.6\%$
  for the fraction of luminosity in a diffuse intragroup component in
  the densest $3^\circ\times 2^\circ$ area of the Leo group, relative
  to the light in galaxies. (iii) Using this Leo field as a blank
  field, we derive an average fraction of 13.6\% background emission
  line objects that enter in surveys of Virgo intracluster PN.  This
  is in agreement with an earlier estimate (15\%) obtained from the Ly
  break galaxy population at $ z = 3.13$.  The small fraction confirms
  the validity of the selection criteria for intracluster PN
  candidates in Virgo.

 }
   
\authorrunning{N. Castro-Rodr\'\i guez et al.}

\titlerunning{Narrow band survey in Leo HI cloud}   

\maketitle


\section{Introduction}

Very little is known about the fraction of the diffuse stellar
component in low density environments, such as in galaxy groups. If
most of the intracluster light (ICL) originated by being removed from 
galaxies through collisions (Moore et al.\  1996), the fraction of intragroup
light (IGL) could be significantly less than that seen in rich
clusters. Numerical simulations of mass stripping in compact groups of
galaxies predict a fraction of $5\%-25\%$, depending on the initial
conditions (G\'omez-Flechoso \& Dom\'{\i}nguez-Tenreiro 2001).  So
far, the only observational evidence for the IGL in the intragroup
region is for the M81 group, and results in less than 3$\%$ of the
total stellar luminosity in the group galaxies (Feldmeier et
al.\ 2002).

Recently, significant samples of photometric intracluster planetary
nebulae (ICPNe) candidates in nearby clusters were obtained (Ciardullo
et al.\ 1998; Feldmeier et al.\ 1998; Feldmeier 2000; Arnaboldi et
al.\ 2002, 2003; Okamura et al.\ 2002; Aguerri et al.\ 2003), based on
their strong emission in the [OIII] $\lambda 5007$ \AA\
line. Subsequent spectroscopic surveys (Freeman et al.\ 2000;
Ciardullo et al.\ 2002b; Arnaboldi et al.\ 2003) confirmed that the
majority of these PN candidates are true ICPNe, based on the presence
of the [OIII] doublet in these spectra.  In the Virgo surveyed fields,
the diffuse light may represent 10-40$\%$ of the total light of the
cluster galaxies (Feldmeier et al.\ 1998; Ferguson, Tanvir \& von
Hippel 1998; Okamura et al.\ 2002; Arnaboldi et al.\ 2003; Aguerri et
al.\ 2003).

In the present work we report on the results of a narrow band imaging
survey for emission line objects in the Leo group. The Leo group
consists of early-type galaxies, among which NGC~3379 (M105) and NGC
3384, and the spiral galaxies NGC~3368 (M96) and NGC~3351 (M95)
(Schneider 1989).  Our surveyed field is centred on the peak
of the intergalactic HI emission, which was discovered by Schneider et
al.\ (1983), and later shown to be part of the HI ring in
rotation around M105 and NGC~3384 (Schneider et al.\ 1989).

The survey in the Leo group was carried out with the same instrumental
set-up as for the Virgo surveys. Emission line objects were identified
following the same procedure adopted for ICPN candidates in Virgo
(Arnaboldi et al.\ 2002; Aguerri et al.\ 2003).  We find that no ICPNe
are present in the Leo field, as we will show in the following
Sections. Thus we can set upper limits for the amount of old stellar
light associated with the HI ring, and for the amount of intragroup
light in Leo. In addition, this field can be used as a blank field, to
derive estimates on the density of background emission line objects,
which would fall in the narrow filter bandwidth used for the detection
of ICPNe in Virgo.

This paper is organized as follows: in Section~2 we present the
photometric observations, and summarize the extraction procedure of
the emission line objects.  The spectroscopic follow-up of two
emission line candidates are presented and discussed in Section~3.
The implications for the luminosity associated with the HI cloud in
Leo, the diffuse intragroup light in the Leo group, and the number
density of Ly$\alpha$ emitters in Virgo-related surveys are discussed
in Section~4.  Conclusions are given in Section~5.

\section{WFI survey in the Leo cloud}

\subsection{Observations}

\begin{figure}
\begin{center}
\mbox{\epsfig{file=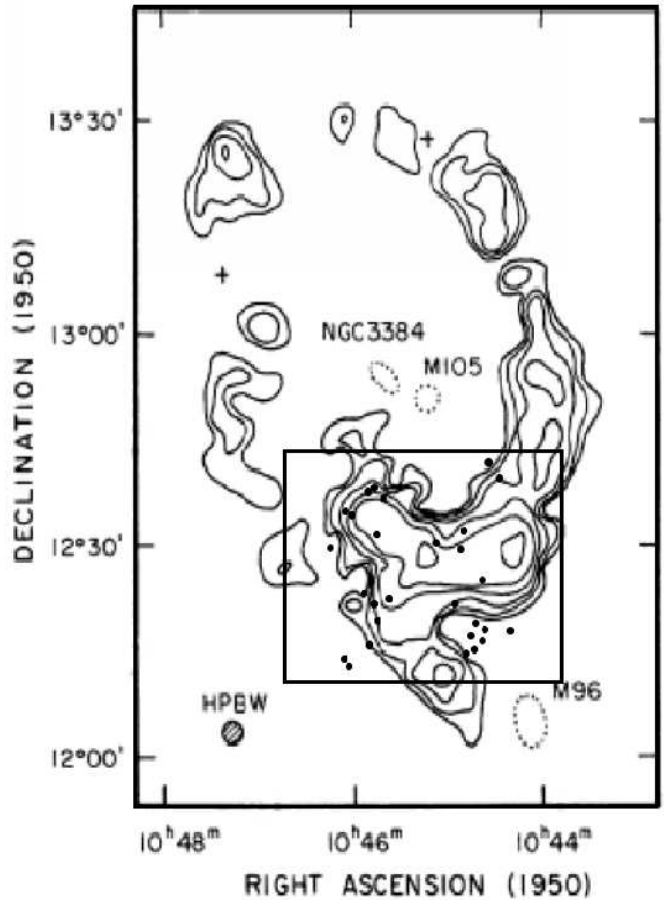,width=10.cm}}
\end{center}
\caption{ Contour plot of the HI cloud in the Leo  Group from Schneider et
al.\ (1989). It  is a ring-like distribution of intergalactic  HI in the M96
group. Contours are shown at 0.1, 0.2, 0.4, 0.8, 1.6  and 3.2 Jy km/s
per beam. 1 Jy km/s per  beam corresponds to a column density of about
2x10$^{19}$ cm$^{-2}$ for the Arecibo beam size. The square represents
the  field  studied in  this  work and  the  filled  circles show  the
positions of the  point-like emission objects with $EW>100$ \AA\ and no  
V-band flux found in this work (see text for details). The nearest
Leo group galaxies are indicated on the plot.}
\label{Fig:fig1}
\end{figure}

We observed a field in the Leo group centered at
$\alpha$(J2000)=10$^{h}$47$^{m}$36$^{s}$ and
$\delta$(J2000)=+12$^\circ\,10'\,56''$ (see Figure~\ref{Fig:fig1}) in
March 1999, during the Observatory of Capodimonte guaranteed time at
the WFI mosaic camera, mounted on the ESO MPI 2.2 meter telescope at
La Silla, Chile.  The camera is mounted at the Cassegrain focus of the
telescope, giving a field of view of $34' \times 33'$.  It consists on
a mosaic of 8 CCD detectors with narrow inter-chip gaps, yielding an
area covering factor of 95.9 \% and a pixel size of $0''.24$. The CCDs
have read-out noise of 4.5 e$^-$ pix$^{-1}$ and a gain of 2.0 e$^-$
ADU$^{-1}$, on average.

To search for planetary nebulae in the intragroup region, we adopted
the ``on-off band" technique.  This technique was developed by
Ciardullo et al.\ (1989a,b) and Jacoby et al.\ (1990) for the
detection of PNe in elliptical galaxies.  Theuns \& Warren (1997),
Arnaboldi et al.\ (2002) and Feldmeier et al.\ (2003) adapted it for
automatic detection on large mosaic images to search for ICPNe in
nearby clusters.  The Leo intragroup field was imaged in the broad
band $V$ filter and in a 28.2 \AA\ wide narrow band filter, centered
at 5026.1 \AA.  The central wavelength and the bandwidth of this
narrow band filter correspond to the redshifted [OIII]$\lambda 5007$
\AA\ emission and the distribution of systemic velocities in the Leo
group. The exposure time for each individual frames was 300 sec in the
V band and 3000 sec in the narrow band [OIII] filter.  A set of 6
frames were taken in each filter, so that the total exposure time is
1800 sec and 18000 sec for the off and on-band, respectively.  These
sequences were acquired in a dithered elongated-rhombi pattern.  This
strategy ensures the removal of CCD gaps in the final co-added image,
a better flat fielding correction, and bad pixel/column removal.  The
atmospheric seeing was $1.2''$ on the final combined images.

\subsection{Data Reduction}

The data reduction was carried out using standard IRAF tasks from the
MSCRED package (see Arnaboldi et al.\  2002 for further details).  The
CCD mosaic frames were de-biassed, dark corrected, and flat-fielded.
Flat-field images were constructed by combining a series of twilight
sky and dome flat images, taken during each observing night.  After
flat-fielding we noticed that some residual structures were still
present in the sky background, which had to be removed before the
final co-added image would be produced.  A ``super-flatfield" image
was constructed from the Leo science images, using a 3 $\times \sigma$
rejection algorithm for the removal of all sources in the field.
The astrometric solution was computed for each image, and then they
were corrected for geometric distortions\footnote{This effect is very
important at the image borders because of their large size.}, sky subtracted,
transparency corrected, and then combined (see Arnaboldi et al.\ 2002).

Landolt fields of standard stars were acquired for flux calibration in the V
band: Landolt 98 and Landolt 107 (Landolt 1992); a spectrophotometric
standard star -- Hiltner 600 (Hamuy et al.\ 1992, 1994) was used for
the calibration of the narrow band filter.  The zero points for the
narrow and broad band filters in the AB magnitude system are:
Z$_{[OIII]}$=20.38 and Z$_{V}$=23.97 for a 1 s exposure.  The
integrated flux from the [OIII] line of a PN is usually expressed in
the $m_{5007}$ magnitude system defined by Jacoby (1989); the AB system
is defined by the flux per unit of frequency (Oke 1990).  Hereafter we
will refer to the [OIII] magnitude in the AB system as $m_{OIII}$. The
relation between these two systems depends on the filter
characteristics (Arnaboldi et al. 2002), and in our case we have:
\begin{equation}
m_{OIII}=m_{5007}-3.54
\label{eq:ec3}
\end{equation}
{ Because the WFI is mounted at the Cassegrain focus of the ESO MPI 2.2m
telescope (f/8), the effects of the converging beam on the trasmission
properties of our interference filter can be neglected.}

\subsection{Catalog of emission line candidate objects}
\label{.cmd}
We have used SExtractor (Bertin \& Arnouts 1996) to carry out the
photometry of all sources in the field.  This program is optimized to
detect, measure and classify sources (star vs. galaxy) on astronomical
images.  It is very useful for the analysis of large extragalactic
surveys. Monte-Carlo simulations were used to set the SExtractor
parameters and optimize source detection.  A background sky image for
the [OIII] image was computed following Arnaboldi et al.\ (2002).  A
population of 1000 point-like objects, with a given luminosity
function (LF), were then added on this image, at random positions.  We
then measured how many objects were detected, the number of spurious
objects due to noise, and the faintest magnitude reached, for a given
detection threshold.  This procedure was repeated several times, with
different SExtractor detection thresholds between 0.7$\sigma$ and
1.3$\sigma$, where $\sigma$ is the count RMS of the background sky
image\footnote{ Detection is performed once the image is convolved
with a Gaussian kernel reproducing the bright star PSF, and at least
5 pixels must be above threshold to trigger detection.}.

        \begin{table*}  \begin{center} \begin{tabular}{cccc}\hline Low
        Threshold & Objects Retrieved & Spurious Objects & Limiting
        Magnitude (m$_{OIII}$)\\ \hline \hline 0.7$\sigma$ & 435 & 404
        & 24.14 \\ 0.8$\sigma$ & 385 & 79 & 24.09 \\ 0.9$\sigma$ & 341
        & 33 & 23.99 \\ 1.0$\sigma$ & 310 & 20 & 23.95 \\ 1.1$\sigma$
        & 277 & 11 & 23.84 \\ 1.2$\sigma$ & 254 & 8 & 23.75 \\
        1.3$\sigma$ & 230 & 7 & 23.67 \\ \hline \end {tabular}
        \end{center}  \caption{List  of  retrieved  objects,  spurious
        objects  detected,  limiting  magnitude for  different
        values of  the detection threshold. The  limiting magnitude is
        the  magnitude  for which  50  \%  of  the input  objects  are
        retrieved.}  \label{Tab:spu} \end {table*}
        
        \begin{figure}                                   \begin{center}
        \mbox{\epsfig{file=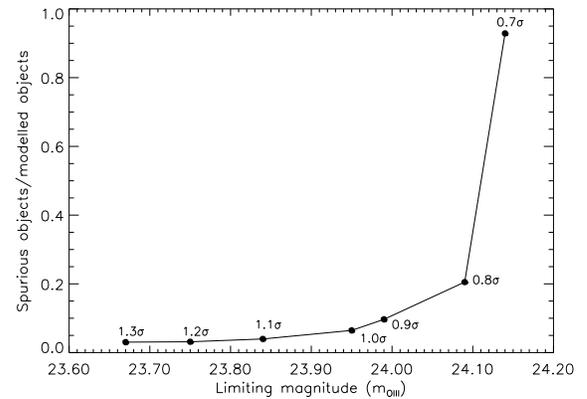,width=8.cm}}          \end{center}
        \caption{Ratios of  spurious to  real detections plotted 
        versus limiting  magnitude for different low thresholds.}
        \label{Fig:fig2} 
\end{figure}
      
\begin{figure*}
\begin{center}
\mbox{\epsfig{file=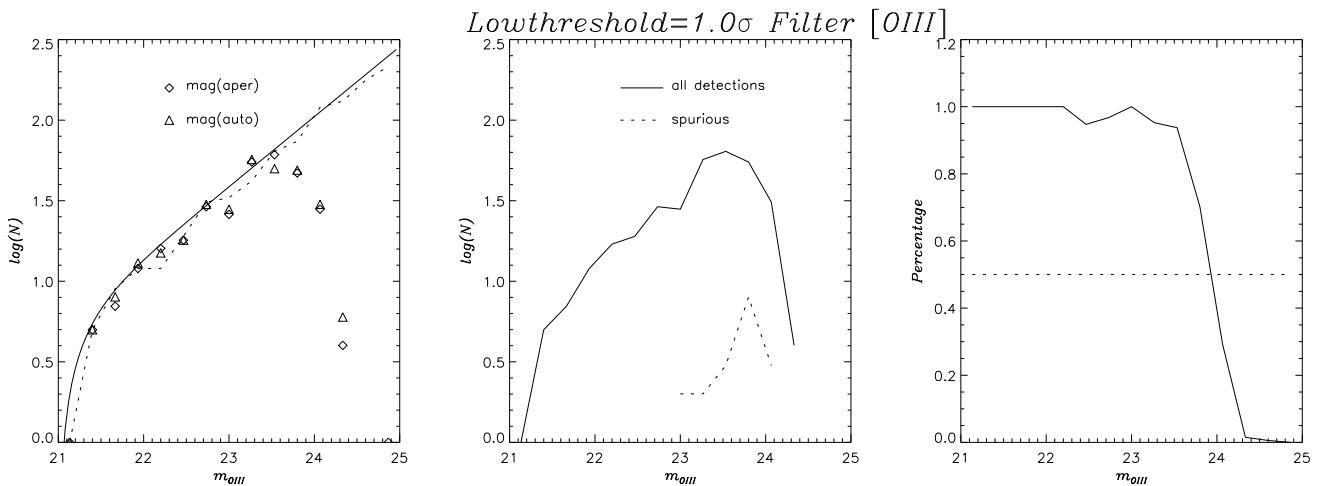,width=18.cm}}
\end{center}
\caption{Monte Carlo simulations of a modeled population of point-like
  objects extracted with SExtractor on the [OIII] image. Left: input
  luminosity function of the modeled objects (continuous line) and
  recovered luminosity function (diamonds and triangles). Center:
  number of detections vs.\ spurious objects for the 1.0$\sigma$
  threshold. Right: fraction of retrieved vs.\ modeled objects; the
  dashed line indicates the 50\% value. The plots are constructed from
  0.26 mag bins.}
\label{Fig:fig3}
\end{figure*}

Table~\ref{Tab:spu} shows the fraction of spurious vs.\ real objects
retrieved in this field, and the corresponding [OIII] limiting
magnitude, for several detection thresholds.  This limiting magnitude
is defined as the magnitude at which 50$\%$ of the input simulated
objects are retrieved.  Figure~\ref{Fig:fig2} shows a plot of the
ratio of spurious vs.\ real objects, for different values of the
detection threshold.  For thresholds lower than 0.9$\sigma$, the
fraction of spurious objects increases rapidly. In the following
analysis we will adopt a conservative threshold of 1.0$\sigma$ for the
source detection in our catalogs.  With this adopted detection
threshold, the limiting magnitude of the [OIII] narrow band image is
23.95 (in the AB system), which corresponds to m$_{5007}$= 27.49.
Figure~\ref{Fig:fig3} shows both the input and recovered LF of the
simulated objects, the spurious and real objects detected, and the
limiting magnitude at the 1.0$\sigma$ threshold.

\begin{figure}
\begin{center}
\mbox{\epsfig{file=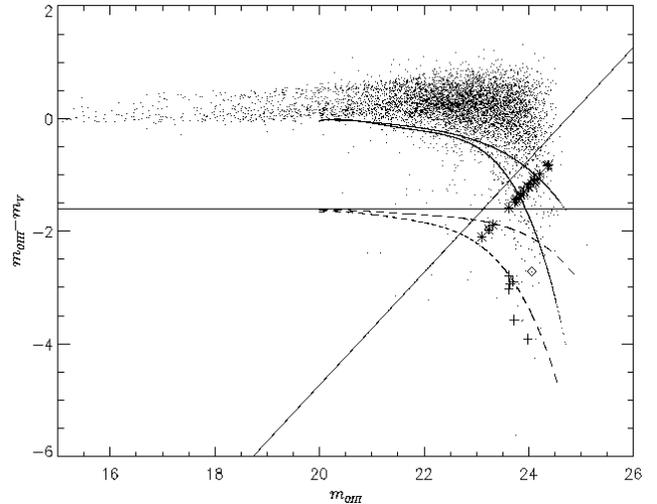,width=9.cm}}
\end{center}
\caption{Color-magnitude diagram (CMD) for all the sources in the Leo field
(dots) detected with SExtractor. The horizontal line represents
the m$_{OIII}$ - m$_{V}$ =-1.6 color, which indicates objects with an
observed EW=100 $\AA$.  The dashed lines represent the 84$\%$ and
97.5$\%$ lines for the distribution of modeled objects with m$_{OIII}$
- m$_{V}$ =-1.6.  Full curved lines represent the 99$\%$ and 99.9 $\%$
lines for the distribution of modeled continuum objects.  The diagonal
line corresponds to m$_{V} =24.73$, which is the 1 $\sigma$ limiting
magnitude in the V band. Crosses indicated the point-like objects
below the color curve of 97.5 $\%$; the asterisks are the objects
without emission in the V band. Because their V magnitude is fainter
than the V limiting magnitude, they do not have a measured position in
this plot ( their y coordinate is unknown); we have adopted a m$_{V}$
= 25.0, to display then on the CMD.  The rhombi
indicate the objects with spectra in this work. See also Arnaboldi et
al.\  (2002) for more detail on the selection procedure.}
\label{Fig:fig4}
\end{figure}

Once the catalog of the [OIII] detected sources is obtained, we
measure the V band flux at the [OIII] source
position. Figure~\ref{Fig:fig4} shows the color-magnitude diagram (CMD)
for the [OIII] detected sources in the Leo intragroup field.
The limiting magnitude in the final V band mosaic image was calculated
using the expression
\begin{equation}
m_{V}=-2.5*\log(4*\pi*\sigma{^2}_{\rm seeing}*{\sigma}_{sky})+Z_V,
\label{eq:ec5}
\end{equation}
where $\sigma_{\rm seeing}$ is the radius of the seeing disk,
$\sigma_{sky}$ represents the RMS of the sky background in the V
band\footnote{ CCD gaps and regions were the background RMS in
the V and [OIII] images were high are masked. The masked total area is
a negligible fraction of the area in the final images.}.  With this
equation, the limiting magnitude is 24.73.

The selection of the most reliable emission line candidates was
carried out following the same criteria as in Arnaboldi et al.
(2002), i.e. these are point-like objects with $EW>100$ \AA\ 
(including the photometric errors), and with broadband emission
fainter than the V limiting magnitude. The final candidates are
indicated with crosses and asterisks in the CMD in
Figure~\ref{Fig:fig4}. A total sample of 29 objects was obtained in the
Leo field.

\begin{table*}  
\begin{center} 
\begin{tabular}{cccccc}\hline

RA (J2000) & DEC (J2000) &  m$_{OIII}$ & err$_{OIII}$ & $m_{5007}$ & $EW_{obs}$ \\
(h m s) & ($^\circ$ ' '') & & & ($\AA$)\\ 
\hline\hline
10:48:41.024 &  12:09:32.27 &  23.10 & 0.17 & 26.55  &  $>100$ \\  
10:48:37.555 &  12:02:40.58 &  23.23 & 0.14 & 26.77  &  $>100$ \\ 
10:48:26.158 &  11:58:38.18 &  23.32 & 0.17 & 26.86  &  $>100$\\
10:47:20.454 &  12:06:39.48 &  23.61 & 0.21 & 27.15  &   509 \\
10:47:00.162 &  12:01:24.95 &  23.61 & 0.19 & 27.15  &   402\\ 
10:48:30.937 &  11:59:23.40 &  23.62 & 0.18 & 27.16  &  $>100$ \\
10:48:26.944 &  12:15:37.48 &  23.64 & 0.23 & 27.18  &   457\\ 
10:47:26.160 &  12:21:44.82 &  23.72 & 0.16 & 27.26  &   847\\
10:48:48.120 &  12:22:40.25 &  23.72 & 0.21 & 27.26  &  $>100$ \\ 
10:48:42.686 &  12:24:36.55 &  23.75 & 0.23 & 27.29  &  $>100$ \\
10:47:30.498 &  12:15:14.04 &  23.77 & 0.19 & 27.31  &  $>100$ \\ 
10:47:05.694 &  11:58:56.78 &  23.82 & 0.23 & 27.36  &  $>100$\\ 
10:48:19.028 &  12:05:15.02 &  23.85 & 0.19 & 27.39  &  $>100$ \\
10:47:24.579 &  12:14:51.43 &  23.88 & 0.20 & 27.42  &   336\\  
10:48:25.801 &  12:13:15.26 &  23.89 & 0.21 & 27.43  &  $>100$ \\
10:48:31.084 &  12:03:31.83 &  23.91 & 0.22 & 27.45  &  $>100$ \\
10:47:07.751 &  12:17:22.55 &  23.91 & 0.23 & 27.45  &  $>100$ \\ 
10:47:25.932 &  12:01:39.90 &  23.98 & 0.21 & 27.52  &   1169 \\
10:47:30.899 &  12:20:52.38 &  23.98 & 0.25 & 27.52  &  $>100$ \\
10:47:29.315 &  12:03:20.04 &  23.99 & 0.24 & 27.53  &  $>100$ \\
10:47:01.458 &  12:13:18.74 &  24.04 & 0.21 & 27.58  &  $>100$ \\
10:47:38.868 &  12:06:27.66 &  24.09 & 0.19 & 27.63  &  $>100$ \\
10:48:30.602 &  12:00:59.27 &  24.10 & 0.21 & 27.64  &  $>100$ \\
10:47:23.065 &  12:21:27.01 &  24.12 & 0.23 & 27.66  &  $>100$ \\
10:47:10.454 &  12:17:30.28 &  24.12 & 0.22 & 27.66  &  $>100$ \\
10:48:52.168 &  12:01:43.69 &  24.21 & 0.27 & 27.66  &  $>100$ \\
10:48:07.852 &  12:14:24.17 &  24.35 & 0.33 & 27.89  &  $>100$ \\
10:47:26.305 &  12:06:12.09 &  24.36 & 0.25 & 27.90  &  $>100$ \\
10:48:36.234 &  12:01:07.16 &  24.38 & 0.26 & 27.92  &  $>100$\\   
\hline  
\end {tabular} 
\end{center}
\caption{Unresolved emission line candidates selected according to the
  criteria established for ICPN candidates in Virgo (Arnaboldi et al.\ 
  2002). The observed equivalent width was computed as $EW_{obs}
  \approx \Delta \lambda (10^{0.4 \Delta m}-1)$, where $\Delta
  \lambda$ is the width of the narrow band filter and $\Delta m$ is
  the object color. Conversion from AB magnitudes to $m_{5007}$ is computed from the filter transmission curve as in Arnaboldi et al. (2002).}
\label{Tab:cand}
\end {table*}

\subsection{Luminosity function and nature of the candidates}
\begin{figure}
\begin{center}
\mbox{\epsfig{file=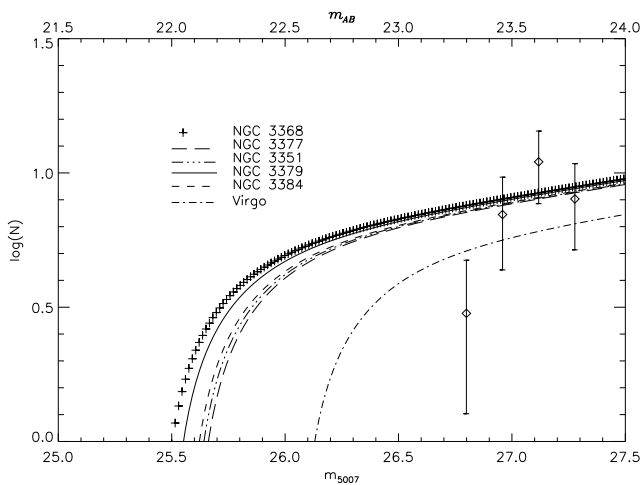,width=9.cm}}
\end{center}
\caption{Luminosity function of the selected emission line candidates 
in the Leo field.
This LF is compared with the PNLF in the Leo galaxies NGC~3379,
NGC~3384, NGC~3377, NGC~3351 and NGC~3368.  The distance modulus
of these galaxies were obtained from the best fit of their PNLF. They
are: 29.96, 30.03, 30.07 (Ciardullo et al.\ 1989a), 30.05 (Ciardullo et al.
\ 2002a) and 29.91 (Feldmeier et al. \ 1997), respectively. The
point-dashed line represents the LF of spectroscopically confirmed
ICPNe in Virgo (Arnaboldi et al.\ 2002)}.
\label{Fig:fig6}
\end{figure}

The PN luminosity function (PNLF) has been used extensively as
distance indicator in late and early-type galaxies (Ciardullo et al.\ 
2002a). Observations in ellipticals, spirals, and irregular galaxies
have shown a PNLF truncated at the bright end.  The shape of this LF
is given by the semi-empirical fit
\begin{equation}
N(M) = c_1 e^{c_2M}[1 - e^{3(M^* - M)}]
\end{equation}
where $c_1$ is a positive constant, $c_2 = 0.307$ and the cutoff
$M^*(5007) = -4.5$ (Ciardullo et al.\ 1989b).  In
Figure~\ref{Fig:fig6}, we plot the PNLF for NGC~3379, NGC~3384,
NGC~3377 (Ciardullo et al.\ 1989a), NGC~3351 (Ciardullo et al.\ 2002b)
and NGC~3368 (Feldmeier et al.\ 1997) which are located in the Leo
group, near the HI cloud, at a distance of 10 Mpc.  The LF of the
spectroscopically confirmed ICPNe located in Virgo (from the FCJ
field; Arnaboldi et al.\ 2002) is also reported in this Figure; its
bright cutoff places the front edge of the Virgo cluster at 12.8
Mpc\footnote{All distances quoted have been determined using the PNLF,
cf.\ Ciardullo et al.\ (1989a).}.  These PNLFs are then compared with
the LF of the selected candidates in the Leo intragroup field. The
bright edge of this LF is $\sim 1.1-1.2$ magnitudes fainter than the
PNLFs in Leo galaxies and $\sim 0.5 $ mag fainter than the PNLF for
the spectroscopically confirmed ICPNe in our Virgo fields. { Both
values are much larger than the photometric errors, see
Table~\ref{Tab:cand}}. Given the well-studied properties of the PNLF
in galaxies, this Figure shows that these emission line candidates are
not a population of PNe at the distance of 10 Mpc, and are not
associated with the HI cloud in the Leo intragroup region.

\subsection{Comparison with the LF of field Ly$\alpha$ emitters}

The bright edge of the LF for the point-like emission line candidates
in the Leo field at $m_{5007} = 26.7$ rules out PNe at a distance of
10 Mpc as a possible explanation for these sources. These objects must
therefore be background emitters, either [OII] emitters at $z\simeq
0.34$, or Ly$\alpha$ emitters at $z\simeq 3.1$.  Our criteria will
select preferentially Ly$\alpha$ galaxies at $z \simeq 3.1$, while
[OII] emitters at $ z\simeq 0.34$ are unlikely because of the EW $ >
100 $\AA\ requirement (Hammer et al.\ 1997; Hogg et al.\ 1998).  We
must then compare the LF of our selected emission line objects with
the LF of Ly$\alpha$ galaxies at $ z = 3.1$.

\begin{figure}
\begin{center}
\mbox{\epsfig{file=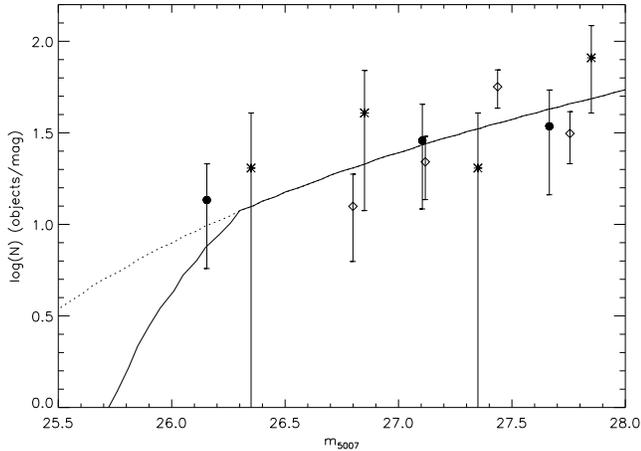,width=9.cm}}
\end{center}
\caption{Luminosity function (LF) of field Ly$\alpha$ emitters at
  $z=3.1$ in our effective surveyed volume. The continuous line is the
  expected LF of the Ly$\alpha$ population in the field at redshift $z
  = 3.13$ from Steidel et al.\ (2000), for which the V magnitude is
  also fainter than 24.73 (our limiting magnitude). The faint dotted
  line shows the expected Ly$\alpha$ LF without considering their V
  magnitudes.  Filled dots indicate the LF of Ly$\alpha$ emitters in
  the blank-field survey by Cowie \& Hu (1998), and asterisks indicate
  the LF of spectroscopically confirmed Ly$\alpha$ emitters from
  Kudritzki et al.\ (2000). Diamonds show the LF of all emission line
  candidates, point-like and resolved, in our Leo field.}
\label{Fig:lfla}
\end{figure}

There is no well-sampled Ly$\alpha$ LF of such objects at $ z = 3.1$
in the literature. Very recent results for the Subaru deep field were
obtained for the Ly$\alpha$ population at $z=4.86$ (Ouchi et al.\
2003), but Arnaboldi et al.\ (2002) derived it from the work by
Steidel et al.\ (2000).  The resulting LF, scaled to our effective
volume here\footnote{The filter FWHM for the Leo survey is a factor
2.83 narrower than for the Virgo survey by Arnaboldi et al.\ (2002).},
is shown as full line in Figure~\ref{Fig:lfla}.  We have also added
data points constructed from the Ly$\alpha$ blank field search done by
Cowie \& Hu (1998), and the spectroscopically confirmed Ly$\alpha$
sample from Kudritzki et al.\ (2000). { Table~\ref{Tab:lf_tab}
gives the data obtained for the different LF plotted on
Figure~\ref{Fig:lfla}}.  The resulting LFs for the field Ly$\alpha$
emitters at $z \sim 3.1$ agree very well with that computed from
Steidel et al.\ (2000).

In these works (Cowie \& Hu 1988; Steidel et al.\ 2000; Kudritzki et
al.\ 2000) Ly$\alpha$ emitters are identified via their narrow line
excess, regardless of whether their emission is point-like or
resolved. Therefore we have produced a catalog of all objects with
line excess ($EW > 100$ \AA), both point-like and extended, to be
compared with the composite field Ly$\alpha$ LF at $z = 3.1$.  The LF
of emission line candidates in the Leo field agrees well with both the
inferred and the observed LF of Ly$\alpha$ emitters at $z= 3.1$;
however, the brightest Ly$\alpha$ emitters with $25.7< m_{5007}< 26.7$
are not present in the surveyed Leo region.

{ From visual inspection of the candidate distribution in
  Figure~\ref{Fig:fig1}, it seems poorly correlated with the HI
  density contours of the Leo gas cloud, and given the properties of
  their LF, which strengthen their association to Ly{$\alpha$}
  emitters at $z\simeq 3.1$, their clustering properties should be
  compared with the results of Ouchi et al. (2003). Unfortunately the
  clustering signal from a sample of 29 objects is very noisy; see
  Aguerri et al. (2003) for additional details.  }

\begin{table}
\begin{center}
\begin{tabular}{ccc}\hline
$m_{5007}$ & log(N) & source\\ 
           & (objects/mag) & \\  
\hline \hline
26.80 & 1.10 & Leo emitters \\
27.12 & 1.34 & \\
27.44 & 1.75 & \\
27.76 & 1.50 & \\
\hline
26.15 & 1.16 & Cowie $\&$ Hu (1998) \\
27.11 & 1.49 & \\
27.66 & 1.57 & \\
28.21 & 1.32 & \\
\hline
26.35 & 1.38 & Kudritzki et al. (2000) \\
26.85 & 1.69 & \\
27.35 & 1.39 & \\
27.85 & 1.99 & \\
\hline
\end {tabular}
\end{center}
\caption{Luminosity functions of the field Ly$\alpha$ emitters in our
effective surveyed volume corresponding to the data plotted in
Figure~\ref{Fig:lfla}.}
\label{Tab:lf_tab}
\end {table}

\section{Spectroscopic follow-up}


\subsection{Observations with FORS2 at UT4-VLT}

The spectroscopic follow-up of two the emission line candidates in the
Leo field was carried out in a subregion, centred at
$\alpha\mbox{(J200)}=10^{h}\,48^{m}\,40.3^{s}$,
$\delta\mbox{(J2000)}=+12^\circ\,01'\,52''$, as a backup program at
UT4 of the VLT at Paranal, on the nights of the 13-14 of April 2002,
with FORS2 in MOS mode.  The FORS2 field of view covers an area of
$6.8' \times 6.8'$ in standard resolution.

Our selected targets were each assigned a MOS slitlet (in MOS mode, up
to 19 movable slitlets can be allocated within the field).
Observations were carried out with GRISM-150I and the order separation
filter GG435+81, giving a wavelength coverage of $4500 - 10200$ \AA\ 
and a dispersion of $6.7$ \AA\ pix$^{-1}$.  The angular scale across
the dispersion is $0''.126$ pix$^{-1}$. The nights were clear, but not
photometric.  The mean seeing was better than $1''.0$ despite strong
northerly winds.  Because the observations were done in blind offset
mode, the slitlet width was chosen to be $1''.4$.  Slitlets were
positioned on the two objects listed in Table~\ref{Tab:can_spec}; IG1
and IG2.  Three stars in the field were selected for pointing checks,
and two stars were used to obtain the correct mask position and
alignment, via the acquisition of their ``through slit'' image.  The
total exposure time was $5\times1800s$.  Spectrophotometric standard
stars were observed at the beginning and end of the nights, but the
conditions made the flux calibration uncertain.

\begin{figure}
\begin{center}
\mbox{\epsfig{file=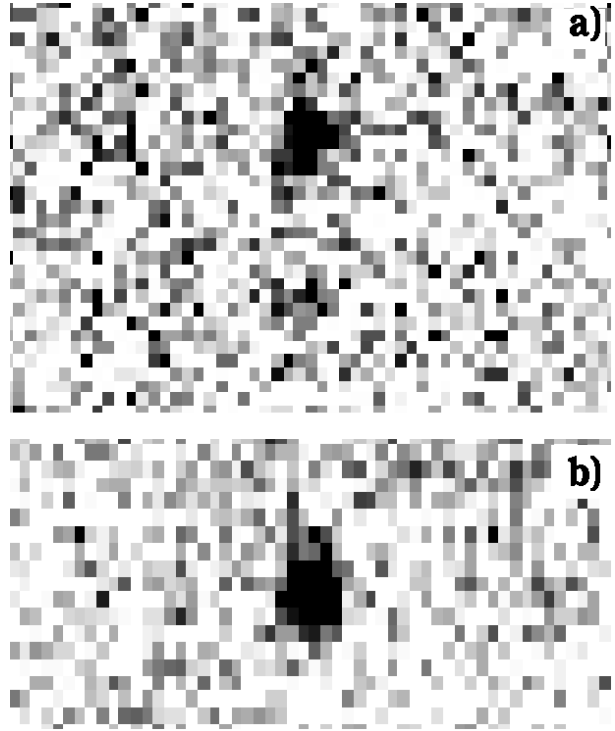,width=8.cm}}
\end{center}
\caption{Two-dimensional (2D) spectra of the emission line candidates
present in the FORS2 field. { Panel a) corresponds object IG2 and panel
b) correspond to object IG1, both listed in Table~\ref{Tab:can_spec}};
gray scale is such that darker = brighter. Each spectrum extends over
400 \AA\ along the x-axis; the upper plot shows $\sim 5''$ along
the y-axis, the lower plot shows $\sim 2''.3$ along the y-axis,
centred on the main emission.  In the upper 2D spectrum there
is a second source in addition to the main unresolved emission IG2, which is
the one selected from our narrow band photometry.  From the wavelength
calibration, the second fainter emission is found at a wavelength
which is in the wing of our narrow band filter bandpass.  The lower 2D
spectrum displays the unresolved single emission from target IG1.}
\label{Fig:fig7}
\end{figure}

\begin{figure}
\begin{center}
\mbox{\epsfig{file=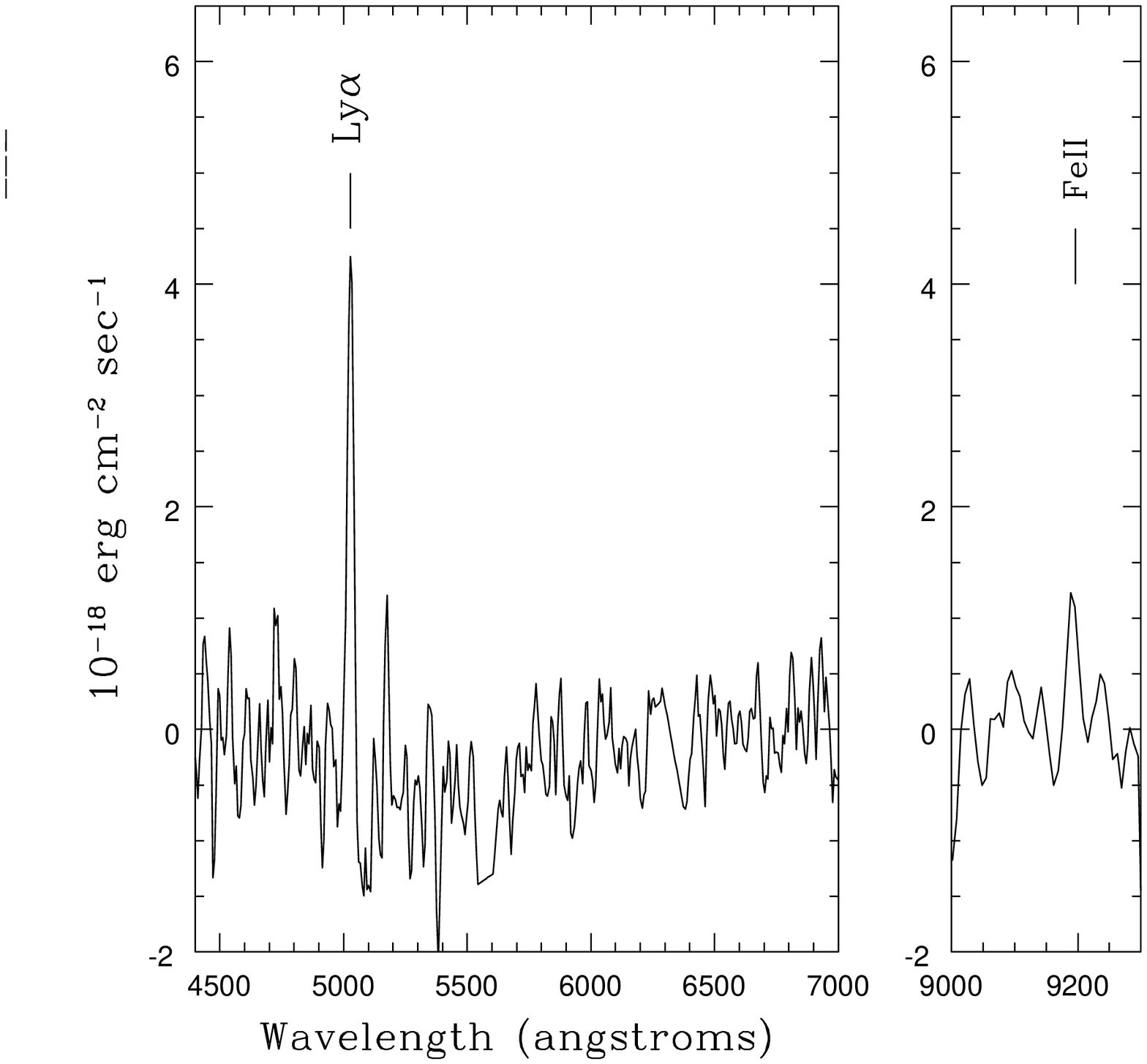,width=8.cm}}
\end{center}
\caption{Spectra in the optical (4500 -7000 \AA) and 
in the NIR (9000 - 9300 \AA) wavelength range for object IG1.
Here the additional emission line in the near-infrared, identified as FeII at
rest frame $\lambda \simeq 2220$ \AA, clearly identifies this object
as a Ly$\alpha$ emitter at redshift $z=3.126$.}
\label{Fig:fig9}
\end{figure}

\begin{figure}
\begin{center}
\mbox{\epsfig{file=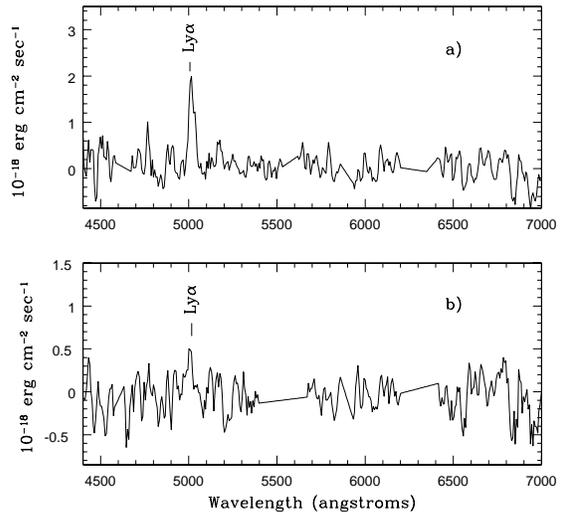,width=8.cm}}
\end{center}
\caption{One-dimensional (1D) spectra of the emission associated with 
target IG2, in the $ \lambda 4500 - 7000$ \AA\ wavelength interval. 
In the wavelength and flux-calibrated 1D spectra, the two spatially
distinct emissions are centred at different wavelengths. 
{\bf a)} Plot of the 1D spectrum associated with the stronger
emission, which we refer to as IG2A, centred at $\lambda 5016.3$ \AA. 
This object is most likely a Ly$\alpha$  emitter at $z=3.12$. 
{\bf  b)} Plot of the 1D spectrum of the second, weaker emission,
which we refer to as IG2B, centred at $ \lambda 5004.6$ {\AA}.  
As discussed in Sect.~3, this is also a Ly$\alpha$ emitter at $z = 3.11$.}
\label{Fig:fig8}
\end{figure}

\begin{table*}
\begin{center}
\begin{tabular}{ccccc}\hline
Name & $\alpha$(J2000) & $\delta$(J2000)&  OIII & V \\ 
     &   (h m s)       & ($^\circ$ ' '')&       &   \\   
\hline \hline
IG1 & 10 48 37.56 & +12 02 40.58 & 23.2304 & 99.0000 \\ 
IG2 & 10 48 47.92 & +12 02 23.74 & 24.0590 & 26.7774 \\  
\hline
\end {tabular}
\end{center}
\caption{List of the  emission line candidates selected for spectroscopic
follow-up, in the subregion centred at
$\alpha\mbox{(J2000)}=10^{h}\,48^{m}\,40.3^{s}$,
$\delta\mbox{(J2000)}=+12^\circ\,01'\,52''$.
}
\label{Tab:can_spec}
\end {table*}

IG1 is the second brightest object listed in Table~\ref{Tab:cand}.
IG2 was not selected as a PN candidate for Table~\ref{Tab:cand},
because of its position in the CMD above the 97.5\% line for
EW$=100\AA$.  Its position in the CMD is caused by errors in the
narrow band photometry, see Figure~\ref{Fig:fig4}.

Two-dimensional (2D) spectra of the selected targets are shown in
Figure~\ref{Fig:fig7}; IG2 turns out to be a multi-component emission,
with an additional fainter source near the strongest emission source
detected in our imaging survey.  1D-spectra for IG1 and IG2,
wavelength and flux-calibrated, are shown in Figure~\ref{Fig:fig9} and
\ref{Fig:fig8}.

In Table \ref{Tab:lines} we report on the parameters determined for
the emission lines detected for IG1, IG2A, and IG2B, via a Gaussian fit.
All detected lines near $\lambda 5020$ \AA\ have a FWHM of several
times the instrumental resolution.

\begin{table*}
\begin{center}
\begin{tabular}{lcc}\hline
Name &  Wavelength & FWHM  \\ & {\AA}  & {\AA}  \\\hline
                        \hline 
                        IG1 (Line 1)  & 5028.5 & 25.0 \\ 
                        IG1 (Line 2)  &  9191.0 &13.0  \\
                        IG2A & 5016.3 &  35.0\\ 
                        IG2B & 5004.6 &  34.0\\ 

\hline

\end {tabular}
\end{center}
\caption{Parameter of the emission lines detected for IG1 and IG2 as
derived from a Gaussian fit to the spectra shown in
Figures~\ref{Fig:fig9}.  and \ref{Fig:fig8} }
\label{Tab:lines}
\end {table*}


\subsection{Results}
In the spectrum associated with IG2A, there is no emission at $\lambda
4965.5$ \AA, which would correspond to the redshifted $\lambda 4959$
\AA\ emission of the [OIII] doublet, had the strongest emission been
the [OIII] $\lambda 5007$ \AA\ PN emission.  We can also exclude that
this strong emission is a redshifted [OII] $\lambda 3727$ \AA\ line of
a starburst emitter at redshift $z= 0.345$, because the redshifted
H$\beta$ and [OIII] $\lambda 5007$ \AA\ emissions should then be
visible in the red part of the spectrum (respectively at $\lambda
6539$ and $\lambda 6736$ \AA), and we can similarly exclude [MgII]
$\lambda 2798$ \AA\ at $z=0.79$.  In addition, the observed line is
broad and asymmetric. This object is most likely a
Ly$\alpha$ emitter at $z=3.12$; see also Kudritzki et al.\ (2000), where
similar findings are discussed for emission line objects in Virgo.

The 1D spectrum of the second (weaker) emission, labeled IG2B, is
centred at $\lambda 5006.2$ {\AA}.  Similar considerations to those
for IG2A suggest that this is also a Ly$\alpha$ emitter, at redshift
$z = 3.11$.

For IG1, again there are no additional lines in the optical, excluding
redshifted [OIII] $\lambda 5007$ \AA, [OII] $\lambda 3727$ \AA, and
[MgII] $\lambda 2798$ \AA\ as identification for the observed emission
at $\lambda 5028.5$ \AA. This, and the fact that the observed line is
broad and asymmetric, argue that we are seeing a Ly$\alpha$ line.  The
near-infrared part of the spectrum shows an additional weak emission
line at $\lambda 9191.0$ \AA. Both observed lines can be explained if
the emission line object is a Ly$\alpha$ emitter at redshift
$z=3.126$, in which case the second line is from red-shifted,
Ly$\alpha$ fluorescent FeII UV emission at $\lambda \simeq 2220$ \AA\
(Sigut \& Pradhan 1998). The other strong lines then expected at
longer wavelengths from fluorescent FeII are not detected in our
spectrum.  They would fall in the region of strong OH sky emission.
{ These FeII UV multiplets are observed in the spectra of active
galactic nuclei (AGN) with broad line regions, and the Ly$\alpha$
fluorescent excitation can more than double the FeII flux in the UV,
producing ultrastrong UV FeII emission (Graham, Cloves \& Campusano
1996). IG1 must then be an AGN at $z\simeq 3.1$ and therefore be
different from the metal-poor, almost dust-free Ly$\alpha$ galaxies
observed by Kudritzki et al.\ (2000).}


\section{Discussion}

The main conclusion of the present work is that there is no population
of PNe at 10 Mpc associated with the intragroup HI cloud in the Leo
group, because (i) the bright edge of luminosity function (LF) for the
emission line objects in this field is $\simeq 1.2$ magnitude fainter
than the bright cut-off of PNLF for the main elliptical galaxies in
Leo; (ii) the LF of the emission line candidates in the field agrees
with the LF for Ly$\alpha$ emitters at $z = 3.1$; and (iii) the
spectra taken for two emission line candidates showed that they are
Ly$\alpha$ objects. None of the acquired spectra showed the typical
[OIII] doublet expected in real PNe.

\subsection{Limits on the surface brightness of the Leo HI cloud}\label{sbleo}
We can set an upper limit to the number of PNe at 10 Mpc,
associated with the HI cloud, of $1\pm1$ PN in the first 2 magnitudes 
fainter than the PNLF bright cut-off:
\begin{itemize}
\item The bright cut-off in the LF for a PN population at 10 Mpc
  would be at $m_{5007} \simeq 25.6$. The brightest object out of our
  list of 29 objects down to $m_{5007} = 28$ has $m_{5007} = 26.6$.
  Given the shape the PNLF, we would expect more than 10 PN brighter
  than $m_{5007} = 26.6$ if the objects we see were PNe. The absence
  of bright PNe is also not a consequence of absorption by dust in the
  HI cloud. With a typical column density of $2\times 10^{19} {\rm
    cm}^{-2}$ and the Galactic relation between absorption $A_V$ and
  column density (Bohlin, Savage \& Drake 1978) we expect a typical
  absorption of $A_V=0.01$ mag.
\item In addition, the comparison of the LF of the emission line
  objects in the Leo field with the Ly$\alpha$ LF at $z =3.1$ and the
  spectroscopic follow-up both indicate that our emission candidates
  are most likely Ly$\alpha$ emitters.
\end{itemize}
Thus we conclude that no PN are detected in the range $ 25.6 <m_{5007}
<28$.

We can use the upper limit of $1\pm1$ PN in the Leo field to set an
upper limit to the luminosity of an old stellar population associated
with the Leo HI cloud. The relation between the two is contained in
the luminosity-specific PN density $\alpha$. This measured quantity
depends on the age and metallicity of the stellar population (Hui et
al.\ 1993). The best estimate for an old bulge population comes from
measurements of the PNLF in M31 (Ciardullo et al.\ 1989b), resulting
in $\alpha_{2B}=28.4 \times 10^{-9} PN L^{-1}_{B \odot}$, where
$\alpha_{2B}$ is the ratio of the number of PN in the first two
magnitudes from the bright cutoff to the B-band luminosity of the
population. From this, we obtain an upper limit to surveyed B-band
luminosity in the field, { which depends on the value assumed for
$\alpha_{2B}$. 

From observations of early-type galaxies, the luminosity-specific PN
density can vary by a factor five (Hui et al.\ 1993), and the number
adopted here is the average value. The upper limit on the surveyed
luminosity is then $L_{B,*} < 1 PN \alpha^{-1}_{2B} = 3.5 \pm
^{5.2}_{2.1} \times 10^{7} L_{\odot B}$. The corresponding surface
brightness limit is $< 4.4 \pm^{6.6}_{2.6} \times 10^{-3} L_{\odot
B}\, {\rm pc}^{-2}$ or $\mu_{B,*} > 32.8 \pm 1.0\, {\rm mag\,
arcsec}^{-2}$. }  Previous estimates of the HI cloud surface
brightness were based on CCD photometry, and provided upper limits in
the optical and NIR band.  Skrutskie, Shure \& Beckwith (1984) gave
upper limits to $\mu_{V,*}=28.0$ mag arcsec$^{-2}$ and
$\mu_{K,*}=22.8$ mag arcsec$^{-2}$. Then Pierce \& Tully (1985)
provided tighter limit in the B band, down to $\mu_{B,*}=30.2$ mag
arcsec$^{-2}$. Adopting this upper limit for the surface brightness in
the B band, we can compute the associated luminosity in our surveyed
field, and compare it with our new estimate.  From Pierce \& Tully
(1985), the total B luminosity in our survey field would be
$L_{B,*}=5.1 \times 10^{8} L_{\odot B}$.  The fact that no PNe is
found 2 magnitudes down the PNLF bright cut-off, implies a stronger
limit on the total luminosity, i.e $L_{B,*} < 3.5 \times 10^{7}
L_{\odot,B}$, which is an order of magnitude fainter than the Pierce
\& Tully (1985) estimates.

The mass estimates of Schneider et al.\ (1983) for the HI cloud can be
combined with our surface brightness upper limit and provide a lower
limit to the $M({\rm HI}) / L_B$ ratio. Our surface brightness limit
at the position of the HI peak correspond to $4.4\times
10^{-3}$ L$_\odot$pc$^{-2}$. The HI surface density at this position
is at least $0.5$ M$_\odot$pc$^{-2}$, which gives a lower limit of
$M({\rm HI}) / L_B > 110$ M$_\odot$/L$_\odot$.  This number is
significantly higher than the $M({\rm HI}) / L_B$ values of dwarf
galaxies, so the Leo HI cloud cannot be interpreted as a tidally
disrupted gas rich dwarf galaxy, and may not have formed stars yet.

\subsection{Upper limits on the diffuse stellar population in the Leo group}
The upper limit on the B-band luminosity in our survey field also
gives a constraint on the fraction of diffuse light in the Leo group.
As described in Section~\ref{sbleo}, we have $L_{B,*} < 3.5\pm
^{5.2}_{2.1}\times 10^7L_\odot$ in the surveyed field, implying that
the surface brightness of the diffuse stellar population in the field
is $ \mu_{B,*} > 32.8\pm 1.0$ mag arcsec$^{-2}$.  We can take this limit
as characteristic for the inner $6$ deg$^{2}$ region of the Leo group
centred on the HI intragroup cloud, as discussed by Schneider
(1989). { There is evidence that ICL is not uniformly distributed
in clusters, but the field-to-field fluctuations related the
non-uniform distribution of ICPNe are smaller than the uncertainty in
the observed $\alpha_{2B}$ values}. The total luminosity from
individual galaxies in this Leo region amounts to $5.14 \times 10^{10}
L_\odot$; if this flux was distributed uniformly on the $6$ deg$^{2}$
area, this would give a surface brightness of $\mu_{B,gal} = 28.3$ mag
arcsec$^{-2}$.  Comparing with the current limit of $ \mu_{B,*}$, we
obtain an upper limit for the fraction of diffuse light in the inner
Leo group of $1.6\pm^{3.4}_{1.0}$\%.

This upper limit for the fraction of diffuse IGL is at least ten times
smaller than recent measurements of the fraction of intracluster light
in the Virgo cluster (Arnaboldi et al.\ 2002, 2003; Ciardullo et al.\
2002b). This is an interesting result because the inner region of the
Leo group consists of predominantly early-type galaxies, which during
their possible merger formation might have lost a fraction of the
progenitor stars.

\subsection{The Leo field as a blank field: density of background sources}

Given the absence of any intragroup PN at 10 Mpc, the Leo field can be
used as a blank control field for computing the number of background
line emitters which fall in the spectral range of the narrow band
filters used for the detection of ICPNe in Virgo.  The instrumentation
and the selection criteria implemented for the Leo survey are the same
as adopted by Arnaboldi et al.\ (2002) and Aguerri et al.\ (2003) for
their study of the ICPN samples in Virgo. Their ICPN samples were
obtained for three Virgo fields, which we shall refer to hereafter as
RCN1, CORE and FCJ. The atmospheric seeing in the final combined
images for all Virgo and Leo fields is $1''.2$. Also, the limiting
magnitude in the off-band V photometry is very similar in all fields.
Area covered, filter characteristics ($\lambda_c$, and FWHM) and
limiting magnitude of the ICPN samples in these fields are summarized
in Table~\ref{Tab:Leocont}.

\begin{table*}
\begin{center}
\begin{tabular}{cccccccc}\hline
Field Id.& Area of Field & Limiting & Filter & N. of Sources &
         Extrapolated N.& Contamination Rate \\ & (arcmin$^2$) &
         Magnitude& $\lambda_c$ and FWHM (\AA) & in Leo Field & of
         Contaminants & (\%) \\ \hline \hline 
         RCN1 & 943 & 26.71 & 5023, 80 & 1 & 3  & 5 \\ 
         CORE & 943 & 27.21 & 5023, 80 & 7 & 20 & 26 \\ 
         FCJ  & 265 & 27.01 & 5027, 44 & 3 & 2  & 10 \\ 
         Average & & & & & & 13.6$\pm
         6.4$\\ \hline
\end {tabular}
\end{center}
\caption{Estimates of the fraction of background emission line
galaxies in the Virgo ICPN survey fields, based on the Leo blank field.
The seeing all three Virgo fields and in the Leo field is $1''.2$.}
\label{Tab:Leocont}
\end {table*}

We can scale the Leo results to the effectively surveyed volumes and
limiting magnitudes of the Virgo fields, and derive the expected
fraction of background emitters in the ICPN samples. From
Table~\ref{Tab:Leocont}, the average fraction is 13.6\%.  Given the
large EW criterium adopted in our Leo and Virgo surveys, the
background line emitters are most likely Ly$\alpha$ emitters at z$\sim
3.13$.  Arnaboldi et al.\ (2002) computed the LF of Ly$\alpha$
emitters in their sample from the work of Steidel et al.\ (2000). The
fraction of the expected contamination by Ly$\alpha$ emitters in the
first magnitude of their ICPN sample was $15\%$.  This value is in
agreement with the results from the Leo blank field.

Previous studies have investigated the fraction of background
emission-line objects in the ICPN samples in Virgo.  Freeman et al.\ 
(2000) provided an estimate of 26\% based on the spectroscopic
follow-up for an ICPN sample from Feldmeier et al.\ (1998).  Ciardullo
et al.\ (2002b) carried out a survey at $\lambda 5019$ \AA\ for
emission-line sources in a ``blank-field'' . Once their results are
scaled to the Feldmeier et al.\ (1998) and Ciardullo et al.\ (1998)
surveys, the reported fraction of high-redshift contaminants in their
ICPN samples is 20\%.

These independent estimates are all consistent with a range of
$5-25\%$ for the fraction of background emitters in the Virgo ICPN
samples in different fields. The mean value appears to be $15\%$;
the range of values might be due to clustering of the background
objects (Ouchi et al.\ 2003).

\section{Conclusions}

We have carried out a wide field (0.26 deg$^2$) survey for emission
line objects at $\lambda 5026$ \AA\ associated with the intragroup HI
cloud in the Leo group. A total sample of 29 unresolved emission line
objects were selected according to the criteria adopted for ICPNe in
Virgo.  The luminosity function (LF) of these point-like sources shows
a bright cut-off which is $\simeq 1.2$ magnitude fainter than the
bright cutoff of the planetary nebulae LF (PNLF) from the galaxies
located in the Leo group. Therefore these emission-line objects cannot
be associate with an intragroup PNe (IGPNe) population at the Leo
cloud distance (10 Mpc).  Most likely they are emission line objects
at high redshift.  The subsequent spectroscopic follow-up of two of
these objects showed that they are Ly$\alpha$ emitters at $z \simeq
3.1$.

The absence of PNe in this field constrain the amount of light
associated with the Leo HI cloud. For an old stellar population, the
upper limit to the B-band surface brightness is $\mu_{B,*} > 32.8\pm
1.0$ mag arcsec$^{-2}$, about a factor of ten in flux fainter than
previous limits.  The corresponding lower limit on the HI mass to
B-band luminosity ratio is $M({\rm HI}) / L_B > 110$
M$_\odot$/L$_\odot$, much larger than for dwarf galaxies. This
indicates that the intergalactic gas in Leo is primordial.

The absence of PNe in the Leo intragroup region, down to $m_{5007} =
28$, also allows us to set an upper limit for the amount of intragroup
light in the Leo group. Our data give $L_{B,*} < 3.5 \pm^{5.2}_{2.1}
\times 10^{7} L_{\odot B}$ in this field.  Assuming that this
upper limit applies to the inner $3^\circ \times 2^\circ$ of the Leo
group, the corresponding upper limit to the ratio of diffuse
intragroup to galaxy light is $< 1.6\pm^{3.4}_{1.0}\%$.

This survey can therefore be used as a blank field survey for high
redshift emitters. We can compute the number of background emission
line objects which lies in the spectral range of the narrow band
filters used for the detection of ICPNe in the Virgo cluster.  When
scaled to the number density of ICPN candidates in the Virgo ICPN
samples, the average fraction of high-z galaxies in these samples is
13.6$\%$, in agreement with the estimate of 15\% by Arnaboldi et al.\ 
(2002) based on the Lyman break galaxy population at $z=3$. Including
estimates from spectroscopic surveys, the fraction of background
emitters in the Virgo ICPN samples in different fields ranges from
$5-25\%$, with a mean value of about $15\%$.  The range of values
might be due to clustering of the background objects (Ouchi et al.\ 
2003).

\begin{acknowledgements}
  The authors wish to thank the referee, Dr. J. Feldmeier, whose
  comments have helped them to strengthen the paper. The authors thank
  the ESO 2.2m telescope team for their help and support during
  observations. M.A.\ and O.G.\ thank R.~Scarpa for efficient help at
  UT4.  N.C.\ wish to thank Francisco Garz\'on for his support and
  Norberto Castro-Rodr\'iguez for his help in the computational part
  of this work. This article has made use of data products from NED
  and the CDS public catalogs. J.A.L.A was supported by Spanish DGC
  (Grant AYA2001-3939). This work has been supported by the {\it
  Schweizerischer Nationalfonds} and the {\it Universidad de La
  Laguna-Cajacanarias} fellowship (Canary Islands, Spain).
      
\end{acknowledgements}
      

\end{document}